\documentclass[12pt,a4paper]{article}%
\usepackage{amsmath,amssymb,amsfonts}
\usepackage{graphicx,epsfig}
\usepackage{color}
\usepackage{amsmath}
\usepackage{amsfonts}
\usepackage{amssymb}
\usepackage{float}
\usepackage{caption2}
\usepackage{graphicx}%
\numberwithin{equation}{section}
\baselineskip 0.6cm \setcaptionmargin{1cm}
\begin{document}
\begin{titlepage}
\begin{flushright}
UCB-PTH-06/04
\end{flushright}
\vskip 1.0cm
\begin{center}
{\Large \bf An alternative NMSSM phenomenology with manifest
perturbative unification} \vskip 1.0cm {\large Riccardo
Barbieri$^a$,\ \   Lawrence J.~Hall$^b$, \ Anastasios Y. Papaioannou$^b$,\\[10pt]
Duccio Pappadopulo$^a$\ \ and\ \   Vyacheslav S.~Rychkov$^a$} \\[1cm]
{\it $^a$ Scuola Normale Superiore and INFN, Piazza dei Cavalieri 7, I-56126 Pisa, Italy} \\[5mm]
{\it $^b$ Department of Physics, University of California, Berkeley, and\\
Theoretical Physics Group, LBNL, Berkeley, CA 94720, USA}\\[5mm]
\vskip 1.0cm \abstract{ Can supersymmetric models with a moderate
stop mass be made consistent with the negative Higgs boson searches
at LEP, while keeping perturbative unification manifest? The NMSSM
achieves this rather easily, but only if extra matter multiplets
filling complete $SU(5)$ representations are present at intermediate
energies. As a concrete example which makes use of this feature, we
give an analytic description of the phenomenology of a constrained
NMSSM close to a Peccei-Quinn symmetry point. The related pseudo-Goldstone boson appears
in decays of the Higgs bosons and possibly of the lightest neutralino, and itself decays into $b\bar{b}$ and $\tau\bar{\tau}$.}
\end{center}
\end{titlepage}



\section{Introduction and motivations}

The absence of any clear signal of the Higgs boson(s) at LEP is a disturbing
fact for the Minimal Supersymmetric Standard Model (MSSM). As well known a
heavy stop could be the explanation. Although possible, however, this weakens
the view that requires supersymmetry to be visible at the LHC, especially
since the top, and the stop, have the strongest coupling to the Higgs boson
system. In the attempt to avoid this quite unpleasant road, several proposals
have in fact been made: among them, the consideration of the Next to Minimal
Supersymmetric Standard Model (NMSSM) has received a great deal of attention.
It is in fact true that the extra contribution to the quartic Higgs coupling
arising in the NMSSM can easily accommodate a lightest Higgs boson even much
heavier than in the MSSM\cite{Harnik:2003rs},\cite{Barbieri:2006bg}. A strong
constraint, however, to which we stick in this paper, is its compatibility
with manifest perturbative unification.

Although the NMSSM is a very minimal extension of the MSSM, it has a drawback:
it allows to introduce several more parameters, which often make the various
analyses difficult to follow or can even obscure the very search for
significant phenomenological patterns. In this paper we try
to clarify a possibility offered by the NMSSM to comply with the LEP
constraints\footnote{For other attempts see e.g.\ \cite{Dermisek:2005ar}%
,\cite{Chang:2005ht},\cite{Dermisek:2007ah}. There are, however, two different
aspects of the problem that may have not been equally addressed in these
works: the level of fine-tuning in the $Z$-mass and the narrowness of the
region of parameter space consistent with current data.} in a weakly
fine-tuned and not too narrow region of its parameter space, while insisting
on a relatively light stop. A key point is that such possibility rests on the
largest possible values of the usual $\lambda SH_{1}H_{2}$ coupling of the
NMSSM consistent with manifest perturbative unification, including the
possible existence of extra matter multiplets filling complete $SU(5)$
representations at intermediate energies\cite{pomarol}.

With CP conserved in the scalar sector, the NMSSM has three CP even and two CP
odd neutral fields. With the standard definition of the Higgs doublets $H_{1}$
and $H_{2}$, the only scalar with tree level coupling to the vector boson
pairs VV, often called $h$ since it is the closest to the Standard Model (SM)
Higgs boson, has the composition\footnote{As usual, $\tan{\beta}=v_{2}/v_{1}$,
$v_{i}=\langle H_{i}^{0}\rangle$,$~v_{1}^{2}+v_{2}^{2}=174$ GeV.}
\begin{equation}
h=h_{1}^{0}\cos{\beta}+h_{2}^{0}\sin{\beta}%
\end{equation}
and tree-level diagonal mass-squared
\begin{equation}
(m_{h}^{0})^{2}=M_{Z}^{2}\cos^{2}{2\beta}+\lambda^{2}v^{2}\sin^{2}{2\beta},
\label{m0}%
\end{equation}
corrected by the well known radiative contribution from top-stop loops (for
moderate mixing, $|A_{t}/m_{\tilde{t}}|\lesssim1$)
\begin{equation}
m_{h}^{2}\simeq(m_{h}^{0})^{2}+\frac{3m_{t}^{4}}{4\pi^{2}v^{2}}\log
{\frac{m_{\tilde{t}}^{2}}{m_{t}^{2}}}. \label{mh}%
\end{equation}
We do not show in this radiative correction a small positive contribution due
to $A_{t}$, because in practice for moderate mixing it is compensated by a
further negative $O(\alpha_{s}\alpha_{t})$ correction at the two-loop order.
Thus the one-loop result (\ref{mh}) remains a reasonable approximation.

Note that equation (\ref{m0}) is valid for any scalar potential of the form
\begin{equation}
V=V^{\text{gauge}}(H_{1},H_{2})+\mu_{1}^{2}(S)|H_{1}|^{2}+\mu_{2}^{2}%
(S)|H_{1}|^{2}-(\mu_{3}^{2}(S)H_{1}H_{2}+h.c.)+\lambda^{2}|H_{1}H_{2}%
|^{2}+V(S)\, \label{pot}%
\end{equation}
i.e., in particular, for any NMSSM which stays perturbative up to the GUT scale.

The phenomenology of the NMSSM in relation with the Higgs boson searches at
LEP certainly depends on the value of $m_{h}$, but crucially also on the
mixings of $h$ with the two other CP even scalars, since $m_{h}^{2}$, being a
diagonal entry of a positive definite squared mass matrix, gives only an upper
bound on the mass squared of the lightest physical CP-even scalar.

With this in mind, this paper consists of two logically independent but also
complementary parts. In the first one we discuss the maximum possible values
of the coupling $\lambda$, and therefore of $m_{h}$, in presence of extra
matter multiplets filling complete $SU(5)$ representations at intermediate
energies (Sect. \ref{maxlambda}). Furthermore, based on the values that we
find for $m_{h}$, we consider a simple and generic 2$\times$2 mixing model
between $h$ and the lightest among the two remaining CP even scalars, which,
before mixing with $h$, do not couple at all with VV (Sect. \ref{mixing}). In
the second part we describe a fully detailed and motivated version of the
NMSSM with an approximate Peccei-Quinn symmetry that realizes the
phenomenological pattern outlined in the first part. This approximate symmetry
restricts the number of effective parameters and makes possible an analytic
description of most of the relevant features we want to underline.

\section{On the maximal value of the $S H_{1} H_{2}$ coupling}

\label{maxlambda}

From eq.~(\ref{m0}) $m_{h}$ is especially sensitive to the value of the
coupling $\lambda$ at the weak scale, which is constrained by demanding that
$\lambda$ stays perturbative in its RGE evolution all the way up to the GUT
scale. More specifically, since $\lambda$ grows with the energy from the weak
to the GUT scale, we require for its value at the GUT scale, $\lambda
_{GUT}<0.3\cdot4\pi$.

\begin{figure}[ptb]
\centering
\includegraphics[width=8cm]{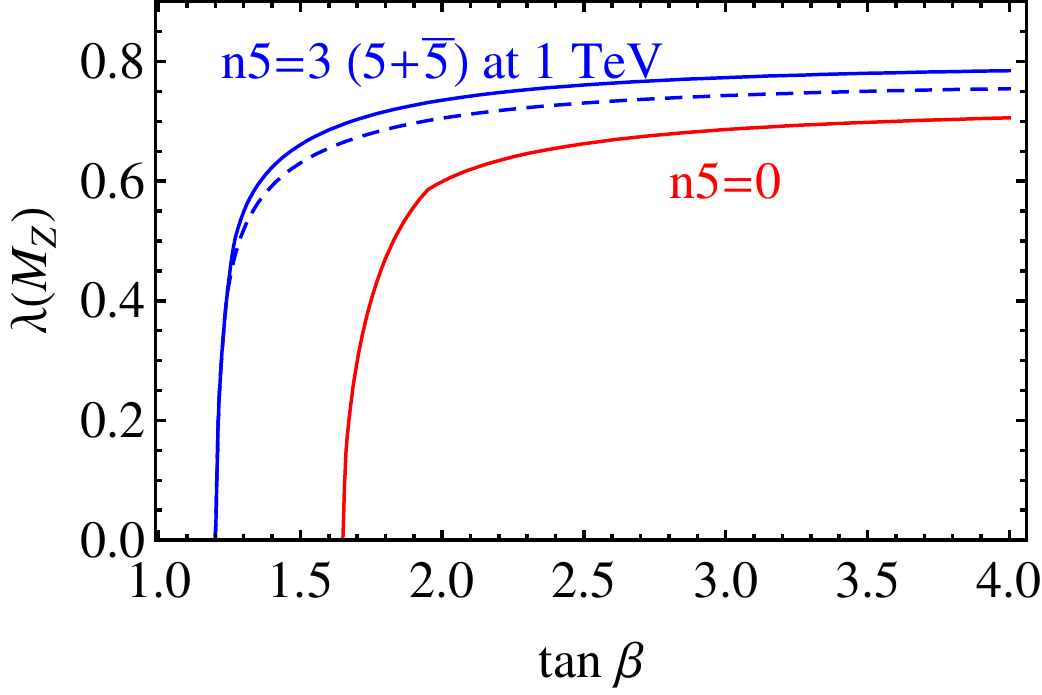}\caption{{\small Lower (red) curve:
the maximal value of $\lambda(M_{Z})$ as a function of $\tan\beta$ in NMSSM
without extra matter at intermediate energies, subject to the condition
$\lambda_{\mathrm{GUT}}/4\pi<0.3$, $\kappa_{\mathrm{GUT}}=0$. Upper (blue)
curves: same but with $n_{5}=3$ extra $5+\bar{5}$ at 1 TeV, and $\lambda
_{\mathrm{GUT}}/4\pi<0.3,0.15$.}}%
\label{lambdamax}%
\end{figure}

\begin{figure}[ptb]
\centering
\includegraphics[width=8cm]{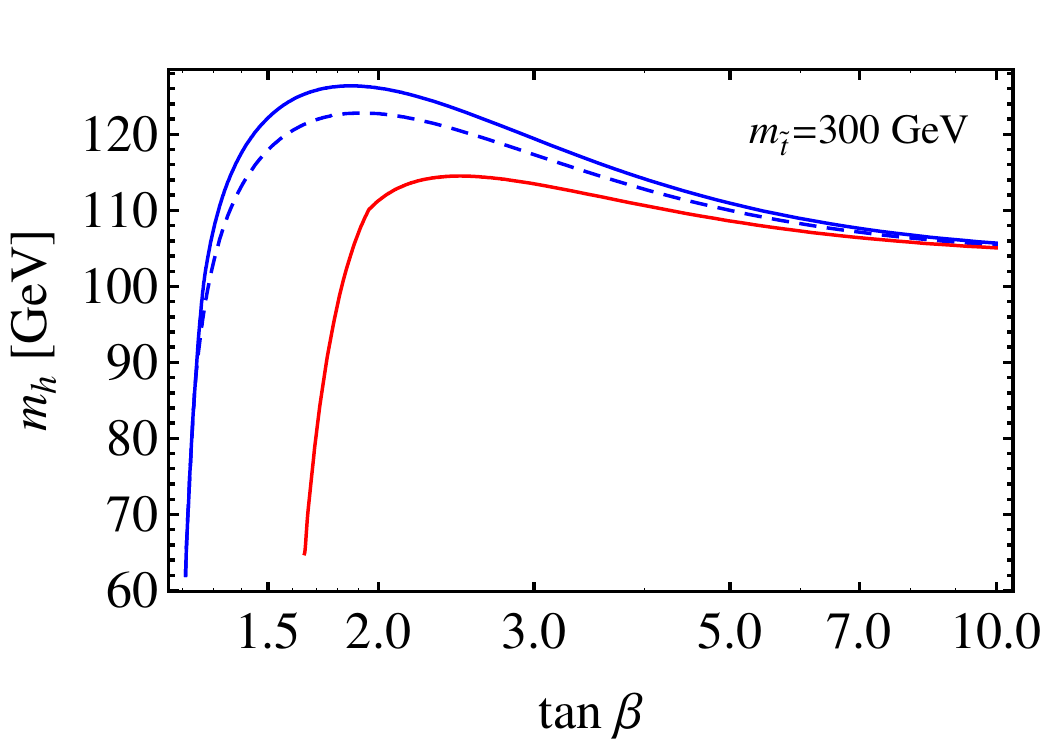}\caption{{\small The maximal value of
$m_{h}$, see Eq. (\ref{mh}), in NMSSM without extra matter (red) and with
$n_{5}=3$ extra $5+\bar{5}$ at 1 TeV (blue). The values of $\lambda
_{\mathrm{GUT}}$ are the same as in Fig.\ \ref{lambdamax}. The stop mass is
fixed at $m_{\tilde{t}}=300$ GeV with moderate mixing, $\vert$}$A_{t}%
/m_{\tilde{t}}|\lesssim1${\small .}}%
\label{mhmax}%
\end{figure}

The RGEs of $\lambda$ and of other relevant couplings can be found in Appendix
\ref{RGEs}. A significant indirect effect on the evolution of $\lambda$ is
caused by the presence at intermediate energies of vector-like supermultiplets
filling complete $SU(5)$ representations \cite{pomarol}. These multiplets
\textit{increase} the gauge couplings at higher energies, which in turn
\textit{slows down} the growth of both $\lambda$ and $y_{t}$, delaying the
onset of nonperturbative behavior. This effect is illustrated in Fig.
\ref{lambdamax} which shows as function of $\tan{\beta}$ the maximum value of
$\lambda$ at the weak scale without or with extra-matter effects (three
$(5+\bar{5})$ of $SU(5)$ at the weak scale), for the current value of
$m_{t}=171$ GeV.\footnote{The example $n_{5}=4$ at $1$ TeV emphasized in
\cite{pomarol} gives non-perturbative values of $\alpha_{\text{GUT}}$ once the
2-loop terms are included in gauge beta functions.} Consequently, from eq.s
(\ref{m0}, \ref{mh}), Fig. \ref{mhmax} gives the maximum value of $m_{h}$ for
a moderate stop mass, $m_{\tilde{t}}=300$ GeV. The upper blue curves, for
$\lambda_{\text{GUT}}/4\pi=0.3,0.15$ are again with three $(5+\bar{5})$ of
$SU(5)$ at the Fermi scale, whereas the lower red curve, for $\lambda
_{\text{GUT}}/4\pi=0.3$, includes in the RGE evolution the standard matter
effects only.

Several features in these figures are worth being observed. All the curves in
Fig. 2 go for large $\tan{\beta}$ to a common asymptotic value which is the
upper bound on $m_{h}$ in the MSSM (for the same $m_{\tilde{t}}=300$ GeV and
moderate mixing). Relative to this value, the increment in $m_{h}$ due to the
extra three $(5+\bar{5})$ is clearly significant, especially since without
extra matter the maximum value of $m_{h}$ barely touches the LEP bound on the
SM Higgs boson mass of about 115 GeV. This is even more so since the upper
limit on $m_{h}$ is essentially saturated for wide variations of
$\lambda_{\text{GUT}}$, in its upper range, as shown by the close upper curves
in Fig. 2.

Both Fig. 1 and 2 are for vanishing $\frac\kappa3 S^{3}$ coupling in the
superpotential, but they are all insensitive to any choice of $\kappa
_{\text{GUT}}/4\pi\leq0.05$, since $\kappa$ is rapidly driven to zero at lower
energies by the RGE evolution. A larger $\kappa_{\text{GUT}}$ would however
reduce the maximum $\lambda$ at the weak scale.

Finally notice that the curves without any extra matter start at $\tan{\beta
}\simeq1.6$ because at lower $\tan{\beta}$, unlike in the case with extra
matter, the top Yukava coupling hits by itself the perturbative bound of
$0.3\cdot4\pi$ before getting to the unification scale.

Recently, Dine, Seiberg and Thomas \cite{seiberg} have claimed that in a
singlet extension of MSSM based on the superpotential $\mu H_{1}H_{2}+\lambda
SH_{1}H_{2}+\frac{1}{2}MS^{2}$ one can raise the Higgs mass by a significant
amount while maintaining manifest perturbative unification \textit{without
extra matter at intermediate scales}. E.g., one of their examples
(\cite{seiberg}, Sec. 4.1) had $m_{h}=120$ GeV for $\tan\beta=4,$
$\lambda=0.7$, $m_{\tilde{t}}=300$ GeV\footnote{In fact they used the soft
stop mass of $300$ GeV, which corresponds to physical stop mass of $345$ GeV,
an irrelevant difference.} and no mixing, which is in clear contradiction with
Fig. 2. As is stressed in the Introduction, our bound on $m_{h}$ applies to
any Higgs potential of the form (\ref{pot}), and in particular to the
superpotential of \cite{seiberg}. We believe that the expansion analysis of
\cite{seiberg}, based on integrating out $S$ and analyzing the spectrum of
light states in terms of coefficients of dimension 6 operators, must be
breaking down, and this explains the discrepancy.

We conclude this Section by analyzing the effect of extra $SU(5)$ multiplets
on the gauge coupling unification.
In Table \ref{unif} we show the prediction of $\alpha_{S}(M_{Z})$ for
$n_{5}=3$ from the running of the gauge couplings at one and two loops,
compared with the standard case ($n_{5}=0$), without any threshold effect. In
the same Table we give, for the two cases, the corresponding value of the
unified coupling $\alpha_{G}$. As is well known, the one loop prediction is
very close to the experimental value $\alpha_{S}(M_{Z})=0.1176(20)$, and of
course this conclusion is left unchanged by the addition of extra matter in
full $SU(5)$ multiplets. At two loops, the prediction for $n_{5}=3$ is brought
closer to the experiment compared to the standard $n_{5}=0$ result. However,
the unavoidable presence of threshold corrections does not allow a significant
distinction between the two cases. In fact, a $i$-th $(5+\bar{5})$ split into
a $SU(3)$-triplet of mass $M_{di}$ and a $SU(2)$ doublet of mass $M_{Li}$,
$\alpha_{S}(M_{Z})$ gives a further one loop threshold correction
\begin{equation}
\frac{\delta\alpha_{S}(M_{Z})}{\alpha_{S}(M_{Z})}\simeq\frac{9\alpha_{S}%
(M_{Z})}{14\pi}\log{\frac{M_{di}}{M_{Li}}}\approx2\%~\log{\frac{M_{di}}%
{M_{Li}}}\ . \label{thres}%
\end{equation}
There is furthermore a two loop contribution from $\lambda$ itself, dominated
by the UV,
\begin{equation}
\frac{\delta\alpha_{S}(M_{Z})}{\alpha_{S}(M_{Z})}\simeq-\frac{9\alpha
_{S}(M_{Z})}{56\pi}\log\left(  {\frac{\lambda_{GUT}^{2}}{2\pi^{2}}\log
{\frac{M_{GUT}}{M_{Z}}}}\right)  ,
\end{equation}
i.e., numerically, $\delta\alpha_{S}(M_{Z})/\alpha_{S}(M_{Z})\simeq
-(1\div2)\%$ for $\lambda_{\text{GUT}}/4\pi=0.3\div0.2$.





\begin{table}[ptb]
\begin{center}%
\begin{tabular}
[c]{|c|c||c|c|l|}\hline
\multicolumn{2}{|c||}{$n_{5}=0$} & \multicolumn{2}{|c|}{$n_{5}=3$} & \\\hline
$\alpha_{S}(M_{Z})$ & $\alpha_{G}$ & $\alpha_{S}(M_{Z})$ & $\alpha_{G}$ &
\\\hline
\textbf{0.117} & 0.041 & \textbf{0.117} & 0.103 & 1-loop\\\hline
\textbf{0.130} & 0.043 & \textbf{0.123} & 0.154 & 2-loop numerical\\\hline
\textbf{0.129} & 0.043 & \textbf{0.122} & 0.143 & 2-loop analytical\\\hline
\end{tabular}
\end{center}
\caption{{\small Prediction for $\alpha_{S}(M_{Z})$ in the standard case
$(n_{5}=0)$ and for $n_{5}=3$ $(5+\bar{5})$ at $1$ TeV. We use one and two
loop gauge beta functions given in Appendix \ref{RGEs} \textit{without the two
loop contributions of }$\lambda$\textit{ and }$y$, which can later be included
perturbatively. The input $\overline{\text{MS}}$ values are $\hat{\alpha
}(M_{Z})^{-1}=127.918$, $\sin^{2}\hat{\theta}_{W}(M_{Z})=0.23122$. We do not
include any threshold corrections. The last line of the table is obtained by
treating the 2-loop terms as perturbative corrections to the 1-loop results,
following the standard method as described e.g.\ in \cite{langacker}.}}%
\label{unif}%
\end{table}

\section{A simple 2$\times$2 mixing model}

\label{mixing}

\begin{figure}[h]
\centering
\includegraphics[width=6cm]{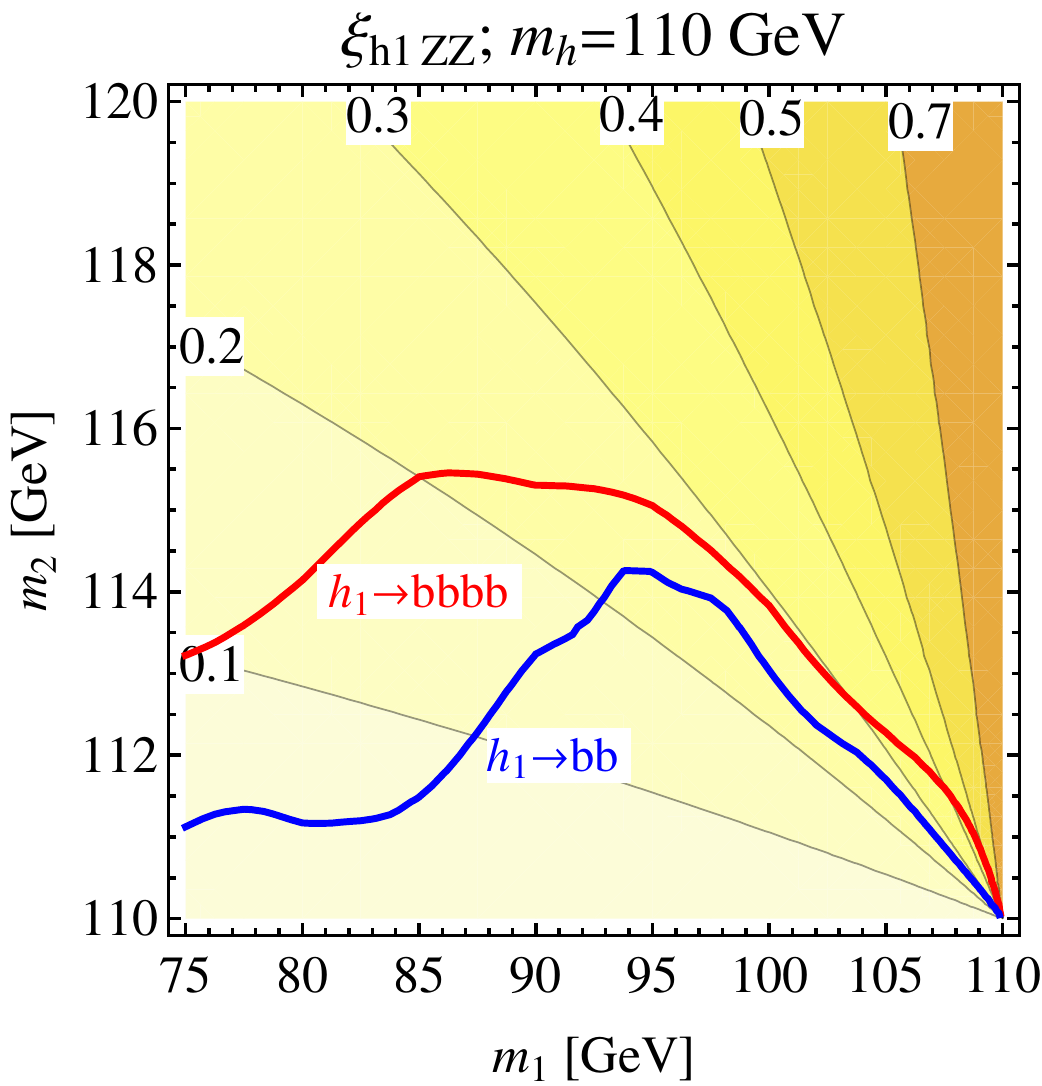}
\includegraphics[width=6cm]{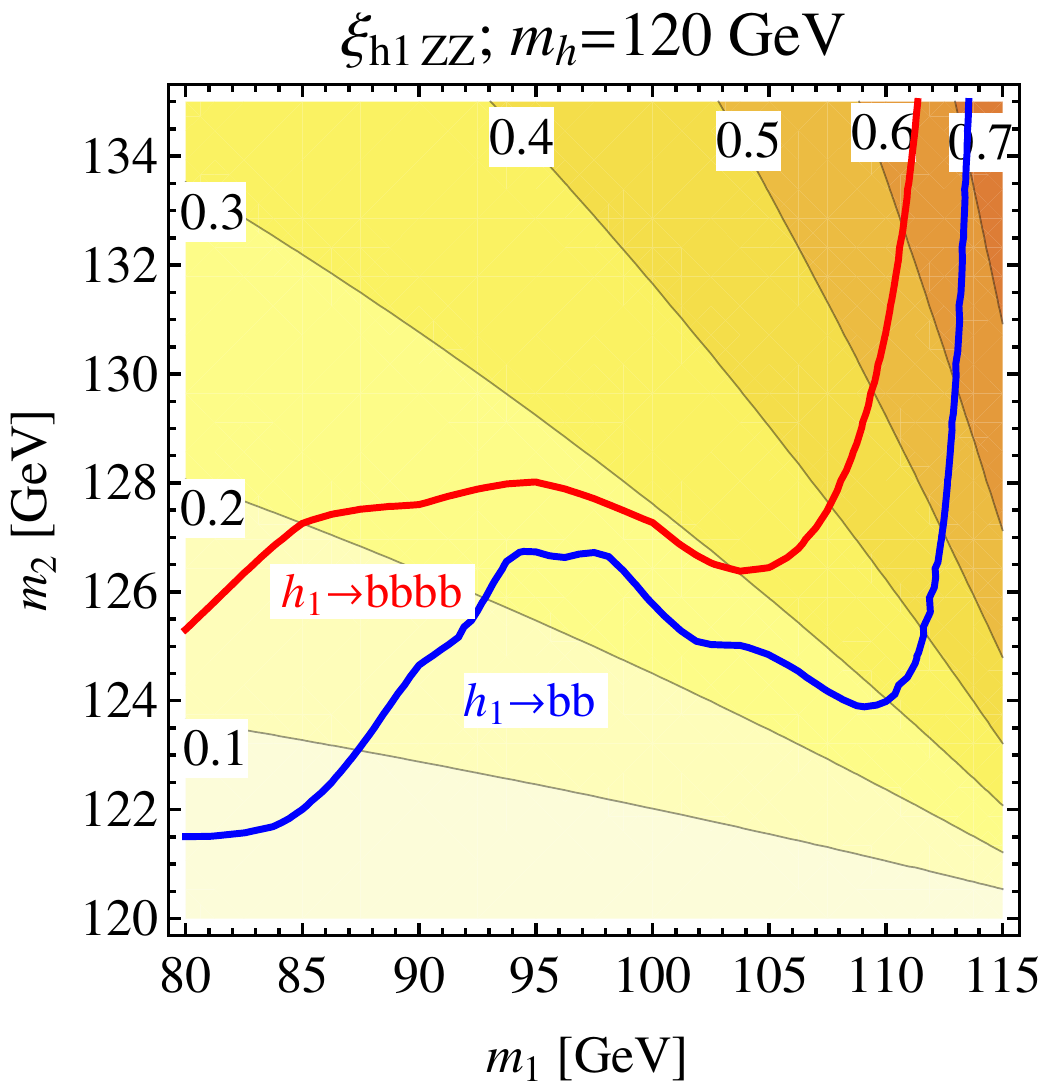}\caption{{\small For two reference
values $m_{h}=110$ GeV (left) and $m_{h}=120$ GeV (right) we plot the
normalized squared coupling, eq. (\ref{xi}), of the lightest scalar $h_{1}$ in
the 2$\times$2 mixing model eq. (\ref{2x2}), plotted as a function of its mass
$m_{1}$ and the mass of the heavier state $m_{2}$. The region below the lower
blue (upper red) curve is consistent with the 95\% C.L. bounds \cite{LEP2}
from nonobservation of $h_{1}$ at LEP2, assuming it decays into $b\bar{b}$
($b\bar{b}\,b\bar{b}$). The heavier scalar $h_{2}$, to be consistent with the
LEP2 searches, should have the mass above $\sim114$ GeV.}}%
\label{m1m2}%
\end{figure}

As already mentioned, what really matters for the NMSSM phenomenology, more
than $m_{h}$ itself, are the masses and compositions of the physical scalars.
Before mixing with $h$, the MSSM\ has two other CP-even fields, $s_{1},s_{2}$
(in their mass-squared diagonal basis). Both their masses and compositions
depend on all the various parameters of the NMSSM. Nevertheless, none of them
is coupled to VV.

Mixing of $h$ with $s_{1}$ and $s_{2}$ (if one, or perhaps both, of these
states are lighter than $h$) can help increase the mass of $h$. After the
mixing, $s_{1}$ and $s_{2}$ acquire coupling to VV and become subject to LEP
searches. As we are going to see, these mixing cannot be large for consistency
with LEP. As such, one can analyze individually their additive effects without
making any significant error. We can then consider a simplified 2$\times$2
mixing model\footnote{The effects of a 2$\times$2 mixing model of this type
has already been considered in the NMSSM \cite{gabe},\cite{Dermisek:2007ah}
and in the MSSM as well \cite{kane},\cite{Drees:2005jg},\cite{Kim:2006mb}%
,\cite{Belyaev:2006rf}.} between $h$ and the lightest, $s_{1}$, among the two
states not coupled to ZZ. Thus we consider a mass matrix%
\begin{equation}
\mathcal{M}_{2\times2}^{2}=\left(
\begin{array}
[c]{cc}%
m_{h}^{2} & \Delta m^{2}\\
\Delta m^{2} & m_{s_{1}}^{2}%
\end{array}
\right)  \label{2x2}%
\end{equation}
with a fixed $m_{h}$ and arbitrary $m_{s_{1}}<m_{h}$ and $\Delta m^{2}$.

In view of the previous Section and having in mind the LEP bound of about 115
GeV, only valid for the SM Higgs boson, we take for $m_{h}$ two reference
values, 110 and 120 GeV, close to the upper bounds on $m_{h}$ without or with
extra matter respectively. In absence of mixing only the latter case would be
compatible with LEP data. With mixing, however, which is generally present,
the situation may change.

In Fig.s 3, we describe the effect of mixing $h$ with $s_{1}$ in the two
cases. In the plane of the two mass eigenvalues $(m_{1},m_{2})$---from which
we can uniquely reconstruct $m_{s_{1}}^{2}$ and $\Delta m^{2}$---we give the
isolines of the squared coupling of the lightest state to ZZ, normalized to
the SM Higgs boson coupling:%
\begin{equation}
\xi_{h_{1}ZZ}=\left(  \frac{g_{h_{1}ZZ}}{g_{hZZ}}\right)  ^{2}. \label{xi}%
\end{equation}
From the data of Ref. \cite{LEP2} this allows to determine in the same plane
the $95\%$ C.L. bound from the non-observation of the lightest state, assumed
to decay in $b\bar{b}$ with SM branching ratio. For later purposes we also
consider the decay in $b\bar{b}$ $b\bar{b}$ with a branching ratio close to 1.
Given the actual numbers, a quick way to understand from these figures the
compatibility with LEP data is to see if there are values of the heaviest mass
$m_{2}$ above 115 GeV and simultaneously allowed by the bound on the lightest state.

The conclusions are quite clear. With an unmixed value of $m_{h}=110$ GeV, and
\textit{a fortiori} for lower values, it is hardly possible to obtain
consistency with the LEP data \footnote{However, notice the point at
$(m_{1},m_{2})\approx(95,115)$ GeV which has been emphasized in the literature
\cite{Drees:2005jg} in connection with a slight excess of events at LEP.}.
This means that, with a moderate stop mass and a small $A_{t}$-term, the NMSSM
\textit{without extra matter} and with standard Higgs boson decays can perhaps
be accommodated with LEP data, if at all, only in a small corner of its
parameter space. This may explain the interest of considering the decay of the
lightest state into $\tau\bar{\tau}$ $\tau\bar{\tau}$, which is experimentally
less constrained \cite{Dermisek:2005ar}.

On the other hand, the $m_{h}=120$ GeV case is obviously compatible with LEP
data for small enough mixing. More important is that some mixing effects will
inevitably be present, which can push the heavier state even further up with a
somewhat reduced coupling to the ZZ, while keeping consistency with the LEP
data for the lower state. This can be a characteristic feature of the NMSSM
with extra matter contributing to the RGE running of the coupling constants,
and is the phenomenological pattern to which we want to draw attention.

\section{An explicit example based on an approximate Peccei-Quinn Symmetry: PQ
SUSY}

\label{PQ}

\subsection{The Lagrangian and the allowed parameter space}

An independent motivation for the NMSSM is that it may provide a simple
solution of the so called $\mu$-problem: the supersymmetric superpotential
mass term $\mu H_{1}H_{2}$ gets replaced by $\lambda\langle S\rangle
H_{1}H_{2}$ and all the mass terms in the Lagrangian originate from
supersymmetry breaking. This possible solution of the $\mu$-problem invites a
symmetry explanation of the absence of mass terms in the superpotential. Such
symmetries can be a continuous R-invariance and/or a Peccei-Quinn (PQ)
symmetry. In this paper we choose a PQ symmetry since: i) it removes the
$\frac{\kappa}{3}S^{3}$ coupling, thereby helping to maximize $\lambda$ at the
weak scale; ii) it can reduce the number of parameters in the supersymmetry
breaking Lagrangian as well, since PQ may be approximately realized in this
sector without conflicting with experiments. This version of NMSSM, which we
call \textquotedblleft PQ SUSY," has a minimal number of parameters, and
contains a light pseudo-Goldstone boson. For earlier considerations of NMSSM
in the PQ limit, see \cite{lawrence},\cite{zerwas}.

Up to the small breaking of the PQ symmetry, the Lagrangian is uniquely fixed
by the superpotential term
\begin{equation}
f=\lambda SH_{1}H_{2}, \label{superpot}%
\end{equation}
by the soft non-supersymmetric piece of the scalar potential
\begin{equation}
V_{\text{soft}}=m_{S}^{2}|S|^{2}+m_{1}^{2}|H_{1}|^{2}+m_{2}^{2}|H_{2}%
|^{2}+(A_{\lambda}\lambda SH_{1}H_{2}+\text{H.c.}), \label{Vsoft}%
\end{equation}
and by the gaugino mass terms, which we shall take large relative to
$\lambda\langle S\rangle$ (see below). Small breaking terms of the PQ
symmetry, like $\delta V=m^{2}S^{2}+B\mu H_{1}H_{2}+$H.c.$\,$, will have to be
present. However we assume them to be small enough only to give mass to the
otherwise massless pseudo-Goldstone boson, without significantly affecting any
of the remaining properties of the model. We have checked that this is a
consistent approximation.

\begin{figure}[ptb]
\centering
\includegraphics[width=8cm]{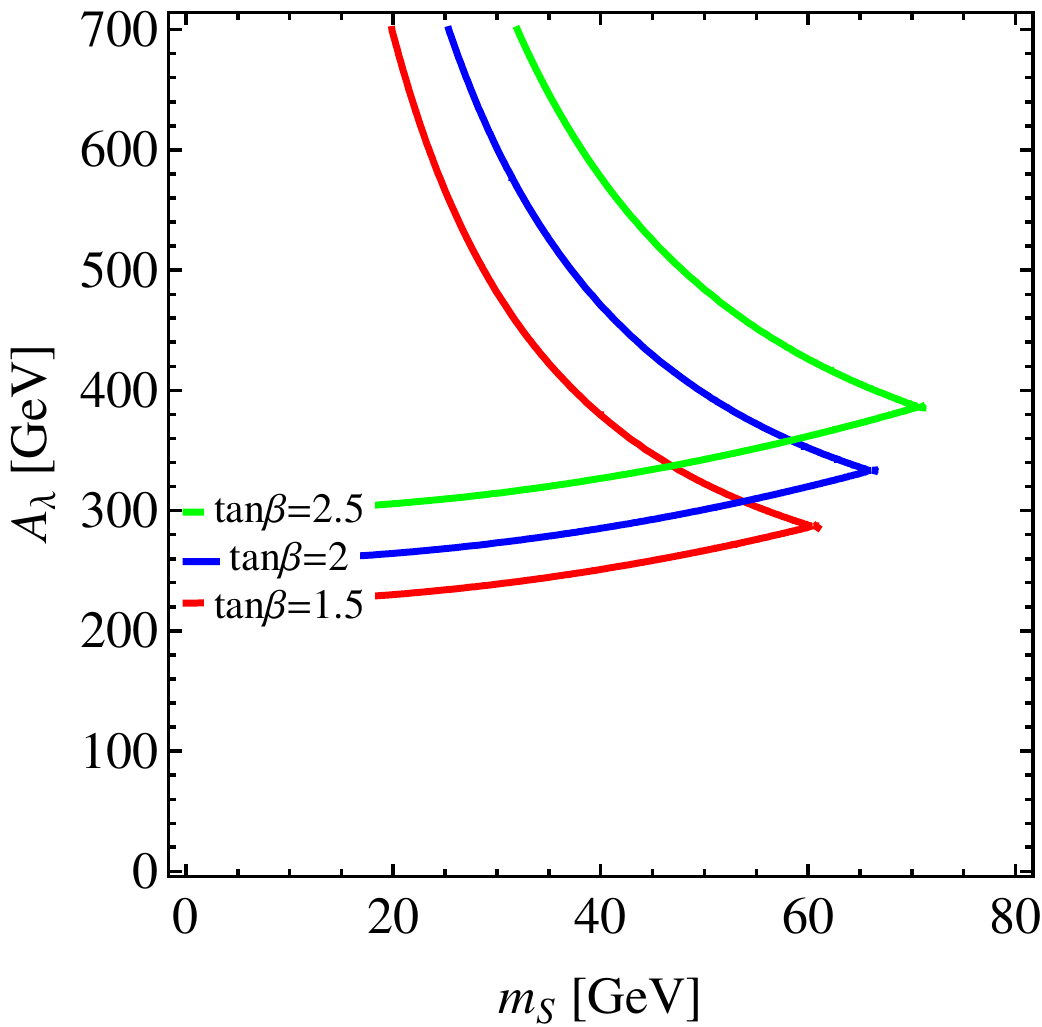}\caption{{\small The allowed region of
the $(m_{S},A_{\lambda})$ plane (see the text) for $\tan\beta=1.5,2,2.5$ and
$\lambda$ fixed at $0.65,0.7,0.75$, respectively.}}%
\label{param}%
\end{figure}

When it exists, the CP-conserving, $SU(2)\times U(1)\rightarrow U(1)$ breaking
vacuum is related to the Lagrangian parameters by $(x=m_{S}^{2}/\lambda
^{2}v^{2})$
\begin{equation}
\lambda^{2}v^{2}=M_{Z}^{2}+\frac{A_{\lambda}^{2}}{1+x}+\frac{m_{1}^{2}%
-m_{2}^{2}}{\cos2\beta}\,, \label{lv}%
\end{equation}%
\begin{equation}
\sin^{2}{2\beta}=2\left[  (1+x)-(1+x)^{2}\frac{m_{1}^{2}+m_{2}^{2}+\lambda
^{2}v^{2}}{A_{\lambda}^{2}}\right]  \,, \label{sin2b}%
\end{equation}%
\begin{equation}
\langle S\rangle\equiv v_{s}=\frac{A_{\lambda}}{2\lambda(1+x)}\sin2\beta\,.
\label{svev}%
\end{equation}
Note that the scalar sector defined by eq.s (\ref{superpot},\ref{Vsoft})
depends upon five parameters (apart from the pseudo-Goldtone mass $m_{G}$):
$\lambda$ and $m_{1}^{2},m_{2}^{2},m_{S}^{2},A_{\lambda}$. We trade
$m_{1,2}^{2}$ for $v$ and $\tan{\beta}$. A useful way to represent the various
results, which we shall follow, is to show them in the plane $(m_{S}%
,A_{\lambda})$ for fixed values of $\lambda,\tan{\beta}$. In particular the
vacuum in eq.s ({\ref{lv}, \ref{sin2b}, \ref{svev}) is indeed the true minimum
of the overall potential only in a portion of this parameter space. Fig.
\ref{param} shows the allowed parameter space for }$\tan\beta=1.5,2,2.5${ and
}$\lambda$ close to the maximal allowed values from Fig. \ref{lambdamax}. We
see that $A_{\lambda}$ has a maximal and minimal allowed value for each
$m_{S}$ in an interval $0<m_{S}<m_{S}^{\max}\simeq70$ GeV. Here $m_{S}^{2}>0$
is required by the global stability. The upper limit on $A_{\lambda}$ comes
from imposing that $V<0$ at the minimum ({\ref{lv}, \ref{sin2b}, \ref{svev}),
so that }it is preferred to the trivial stationary point at $v_{1}=v_{2}%
=v_{s}=0$. The lower limit on $A_{\lambda}$ comes from the experimental bound
on the chargino mass, $m(\chi^{\pm})>103$ GeV \cite{neutralino-search}, via
Eq. (\ref{mueff}) below. The constraint of local stability does not further
restrict the parameter space.

\subsection{Higgs boson and higgsino spectra}

\begin{figure}[ptb]
\centering
\includegraphics[width=8cm]{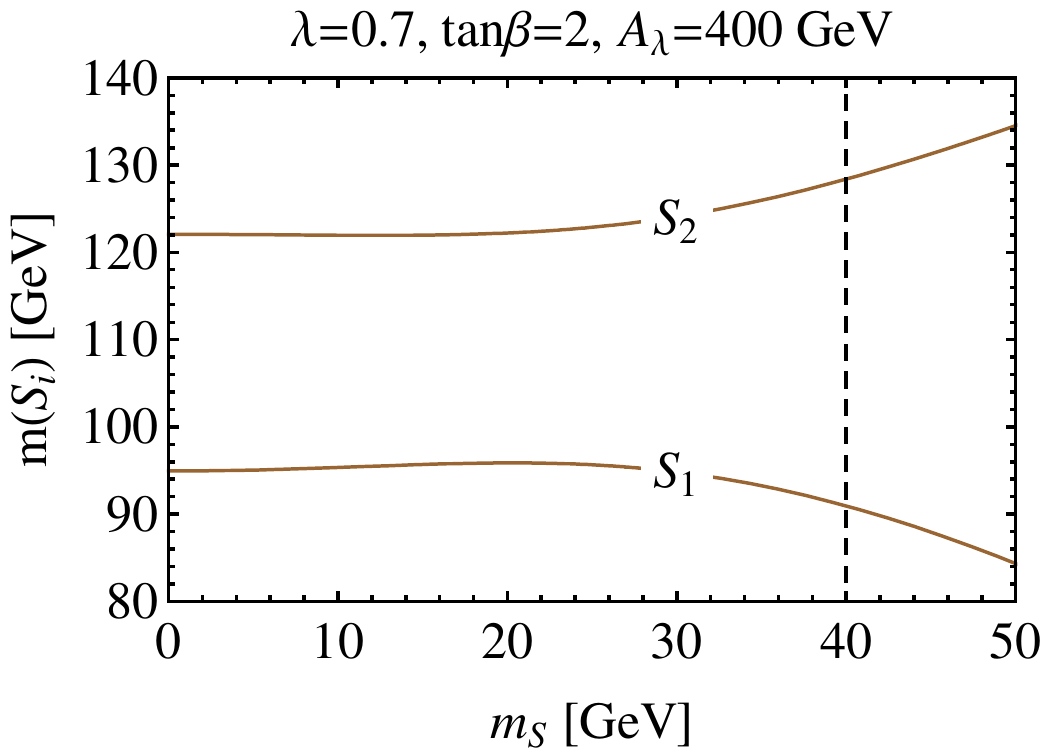}\caption{{\small The masses of the two
lightest CP-even scalars in the model of Section \ref{PQ} for $(\lambda
,\tan{\beta},A_{\lambda})=(0.7,2,400~\mathrm{{GeV})}$ and a range of $m_{S}$
values consistent with global stability of the scalar potential. We include
the stop quantum correction with $m_{\tilde{t}}=300$ GeV. The region to the
right of the dashed line ($m_{S}>40$ GeV) is excluded by LEP constraints on
the $S_{1}ZZ$ coupling (see Fig.\ \ref{paramfinal}). }}%
\label{lightest}%
\end{figure}

\begin{figure}[ptbptb]
\centering\includegraphics[width=8cm]{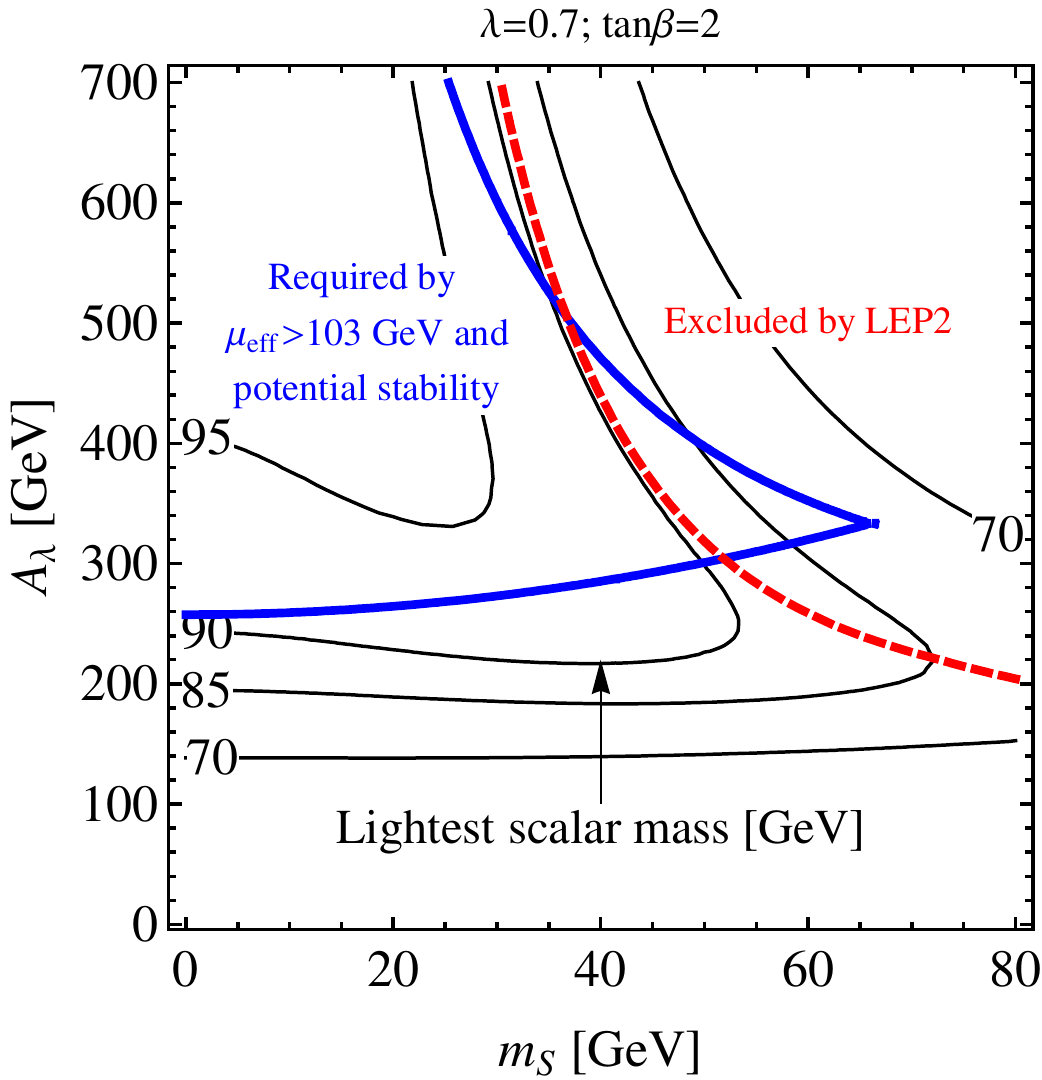}
\caption{{\small For
$(\lambda,\tan\beta)=(0.7,2)$ we give the allowed regions in $(m_{S}%
,A_{\lambda})$ plane following from the stability of the potential and the
chargino mass bound (inside the blue curve, same as in Fig. \ref{param}), and
from the non-observation of $S_{1}\rightarrow GG\rightarrow4b$ decays at LEP2
(to the left of the dashed red curve). In the same plot we show contours of
the lightest CP-even scalar mass. As always, we include the stop quantum
correction with $m_{\tilde{t}}=300$ GeV.}}%
\label{paramfinal}%
\end{figure}

The spectrum of the Higgs boson sector is straightforwardly obtained by
expanding around the above minimum. For the single charged boson one finds
\begin{equation}
m_{H^{\pm}}^{2}=M_{W}^{2}+\frac{A_{\lambda}^{2}}{1+x}-\lambda^{2}v^{2}.
\label{mHpm}%
\end{equation}
Out of the two neutral CP-odd states, one is massless in this approximation
(the PQ Goldstone $G$) and the other has mass
\begin{equation}
m_{A}^{2}=\frac{A_{\lambda}^{2}}{1+x}+\lambda^{2}v^{2}(1+x). \label{mA}%
\end{equation}
Their compositions in terms of the neutral fields
\begin{equation}
H_{1}^{0}=\frac{1}{\sqrt{2}}(h_{1}+i\pi_{1}),~H_{2}^{0}=\frac{1}{\sqrt{2}%
}(h_{2}+i\pi_{2}),~S=\frac{1}{\sqrt{2}}(s+i\pi_{s}),
\end{equation}
are
\begin{align}
G  &  =-\sin\alpha\pi_{s}+\cos\alpha(\cos\beta\pi_{2}-\sin\beta\pi
_{1}),\label{Acomp}\\
A  &  =\cos\alpha\pi_{s}+\sin\alpha(\cos\beta\pi_{2}-\sin\beta\pi_{1}),
\end{align}
where
\begin{equation}
\tan\alpha=\frac{A_{\lambda}}{\lambda v(1+x)}. \label{CP-mix}%
\end{equation}
The $3\times3$ squared mass matrix of the CP-even neutral scalars is best
written in the basis
\begin{equation}
(H=\cos{\beta}h_{2}-\sin{\beta}h_{1},\,h=\cos{\beta}h_{1}+\sin{\beta}%
h_{2},\,s)
\end{equation}
where it has the form
\begin{equation}
\mathcal{M}^{2}=%
\begin{pmatrix}
\frac{A_{\lambda}^{2}}{1+x}+(M_{Z}^{2}-\lambda^{2}v^{2})\sin^{2}2\beta &
-\frac{1}{2}(M_{Z}^{2}-\lambda^{2}v^{2})\sin4\beta & -A_{\lambda}\lambda
v\cos2\beta\\
\ast & M_{Z}^{2}\cos^{2}2\beta+\lambda^{2}v^{2}\sin^{2}2\beta & -A_{\lambda
}\lambda v\sin2\beta\frac{x}{1+x}\\
\ast & \ast & \lambda^{2}v^{2}(1+x)
\end{pmatrix}
\,. \label{mass-matrix}%
\end{equation}
Note, as anticipated, that $\mathcal{M}_{22}^{2}=(m_{h}^{0})^{2}$. Note also
that one of the mixing terms between $h$ and the two other scalars,
$\mathcal{M}_{12}^{2}$ is always small, whereas the other, $\mathcal{M}%
_{23}^{2}$, is essentially controlled by $x$ (or $m_{S}^{2}$). The mixing
pattern discussed in the previous Section is precisely realized in this case,
as shown in Fig. \ref{lightest}, where one has the two lightest scalar masses
as functions of $m_{S}$ for $(\lambda,\tan{\beta},A_{\lambda})=(0.7,2,400~$%
GeV$)$. The lightest scalar mass is below the LEP limit; its dominant decay
mode (see the next Section) is into 2 PQ pseudo-Goldstones: $S_{1}\rightarrow
GG\rightarrow4b$. From Fig. \ref{paramfinal} we see that for $(\lambda
,\tan{\beta})=(0.7,2)$ the LEP constraint on the $S_{1}$ coupling to ZZ is
satisfied in most of the parameter space allowed by the potential stability
and the chargino mass bound\footnote{The processes $e^{+}e^{-}\rightarrow
Z^{\ast}\rightarrow S_{1,2}G$ followed by $S_{1,2}\rightarrow GG$ could not
possibly be seen at LEP2: the normalized squared couplings $\xi_{ZS_{1,2}G}$
\ are tiny, $\lesssim10^{-2}$, one order of magnitude below the LEP2 limits
\cite{LEP2}.}. The heaviest CP-even scalar has mass $m_{S_{3}}\approx
A_{\lambda}/(1+x)^{1/2}$ which for $A_{\lambda}=400$ GeV is in the
$380\div400$ GeV range, while $A$ and $H^{\pm}$ are $\sim15$ GeV heavier and
lighter than $S_{3}$, respectively.



\begin{figure}[ptb]
\centering
\includegraphics[width=8cm]{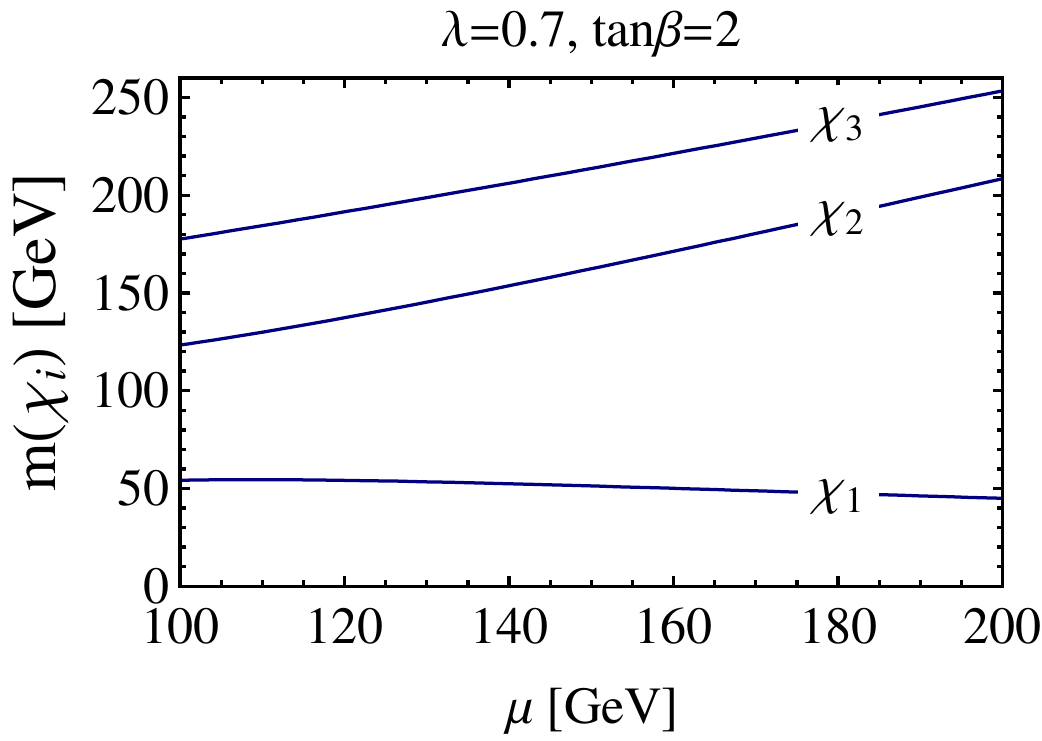}\caption{{\small The neutralino
masses in the model of Section \ref{PQ} for $(\lambda,\tan{\beta})=(0.7,2)$
and the chargino mass $\mu$ in its typical range. We have $m(\chi_{1}%
)+m(\chi_{2})=m(\chi_{3})$ since the neutralino mass matrix (\ref{nmass}) is
traceless. }}%
\label{neutralini}%
\end{figure}

For heavy gaugino masses, the masses of the higgsinos are controlled by the
effective $\mu$-parameter
\begin{equation}
\mu=\lambda v_{s}=\frac{A_{\lambda}}{2(1+x)}\sin2\beta. \label{mueff}%
\end{equation}
The single charged higgsino has mass $m(\chi^{\pm})=\mu$, whereas the
$3\times3$ neutralino mass matrix in the basis $(\tilde{h}_{1},\tilde{h}%
_{2},\tilde{s}$) has the form
\begin{equation}
\mathcal{M}_{\chi}=%
\begin{pmatrix}
0 & \mu & \lambda v\sin\beta\\
\ast & 0 & \lambda v\cos\beta\\
\ast & \ast & 0
\end{pmatrix}
. \label{nmass}%
\end{equation}
Fig. \ref{neutralini} shows the values of the neutralino masses versus
$m(\chi^{\pm})$ in its typical range, $100\div200$GeV, for $(\lambda
,\tan{\beta})=(0.7,2)$. For these masses the LEP2 searches have not been
possibly effective. Indeed, the process $e^{+}e^{-}\rightarrow Z^{\ast
}\rightarrow\chi_{1}\chi_{2}$ is within the LEP2 kinematic limit in a part of
the parameter space (see Fig. \ref{neutralini}). However, the production cross
section turns out to be well below the $\sim0.1$ pb limit set in
\cite{neutralino-search} due to phase space and coupling suppressions.

Another possible process is $e^{+}e^{-}\rightarrow\chi_{1}\chi_{1}%
\gamma_{\text{ISR}}$ with a photon (from Initial State Radiation) and missing
transverse energy in the final state, which is constrained by LEP2 searches of
extra neutrino species \cite{LEP2-photon}. However, we concluded that the
existing data cannot rule out a $\chi_{1}$ with a somewhat reduced $Z\chi
_{1}\chi_{1}$ coupling and mass above $m_{Z}/2$, as it is in our case (see
also \cite{dreiner}).

If the gravitino is the lightest SUSY particle, the lightest neutralino will
predominantly decay into the gravitino and the pseudo Goldstone boson $G$.
However, if the SUSY breaking scale $\sqrt{F}$ exceeds about 1000 TeV, these
decays happen outside the detector and do not modify collider phenomenology of
the model (see Section \ref{conclusions}).

\subsection{Higgs boson couplings and branching ratios}

The phenomenology of the model is made peculiar by the presence of the light
pseudoscalar $G$, with an unknown mass, $m_{G}$, coming from the breaking of
the PQ symmetry and assumed to be relatively small.

If $m_{G}$ is below the $b\bar{b}$ threshold, the pseudoscalar can be seen in
radiative $\Upsilon$ decays \cite{derm-decay}. The relevant branching ratio is
given by \cite{wilczek}
\begin{align}
BR(\Upsilon\rightarrow\gamma G)  &  =F(\cos\alpha)^{2}(\tan\beta)^{2}%
\frac{G_{F}m_{b}^{2}}{\sqrt{2}\pi\alpha_{\text{EM}}}BR(\Upsilon\rightarrow
\mu^{+}\mu^{-})\nonumber\\
&  \simeq2\times10^{-4}\,F(\cos\alpha)^{2}(\tan\beta)^{2},\quad F=\left(
1-\frac{m_{G}^{2}}{m_{\Upsilon}^{2}}\right)  F_{\text{0}},\nonumber
\end{align}
where $\alpha$ is the angle in (\ref{CP-mix}), and the suppression factor
$F_{\text{0}}\lesssim0.5$ is due to QCD, bound state and relativistic
corrections (see \cite{kane-book}, Section 3.1). For ($\lambda,\tan
\beta,A_{\lambda})=(0.7,2,400$ GeV) we get $BR(\Upsilon\rightarrow\gamma
G)\simeq0.5\times10^{-5}.$ The experimental limits on this branching ratio
depends crucially on the decay properties of $G.$ An interesting possibility
occurs if $2m_{\tau}<m_{G}<2m_{b},$ so that $G$ decays into $\tau^{+}\tau
^{-}.$ In this case the current limit from CLEO is \cite{CLEO}%
\[
BR(\Upsilon\rightarrow\gamma G(\rightarrow\tau\tau))\lesssim10^{-4}\text{,}%
\]
and a dedicated run by BABAR may improve it soon by $1-2$ orders of magnitude.

Below we will assume that $m_{G}$ is above the $b\bar{b}$ threshold,
corresponding to a relatively less restricted region of parameter space. The
pseudoscalar then decays into $b\bar{b}$ and $\tau\bar{\tau}$ with branching
ratios close to the branching ratios of the SM Higgs boson.

\begin{figure}[ptb]
\centering
\includegraphics[width=8cm]{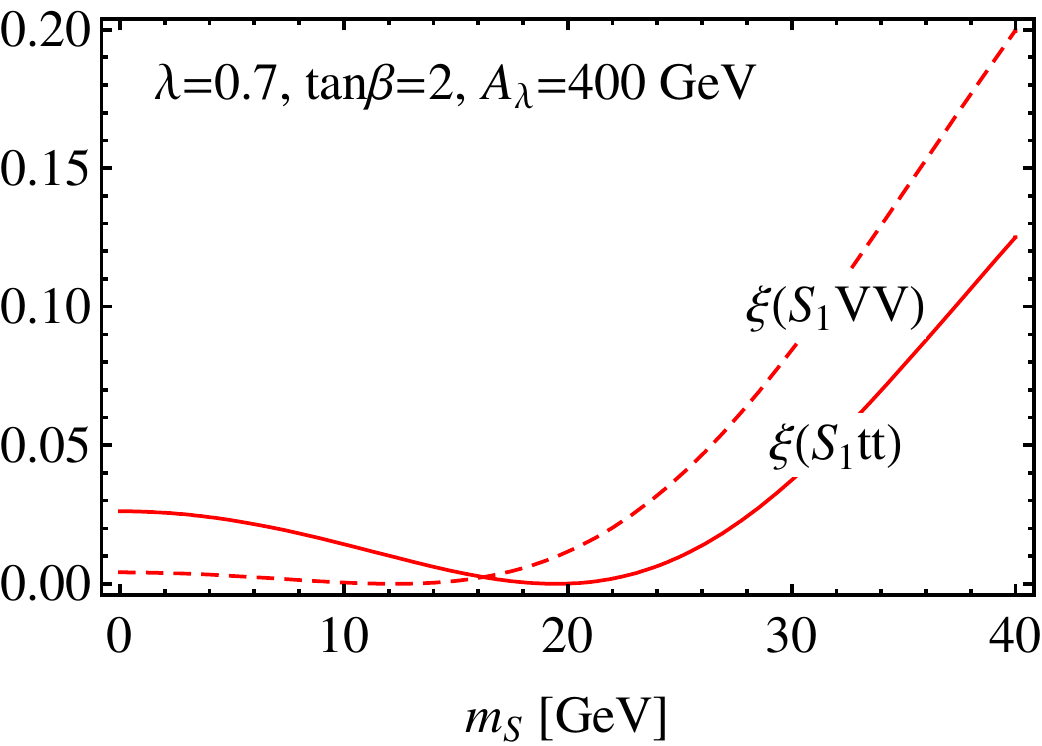}\caption{{\small The $S_{1}t\bar{t}$
and $S_{1}V_{\mu}V^{\mu}$ couplings squared of the lightest CP-even scalar
$S_{1}$, normalized to the couplings of the SM Higgs boson analogously to
Eq.\ (\ref{xi}). In this plot $(\lambda,\tan{\beta},A_{\lambda}%
)=(0.7,2,400\mathrm{GeV})$.}}%
\label{couplS1}%
\end{figure}

\begin{figure}[ptb]
\centering
\includegraphics[width=7cm]{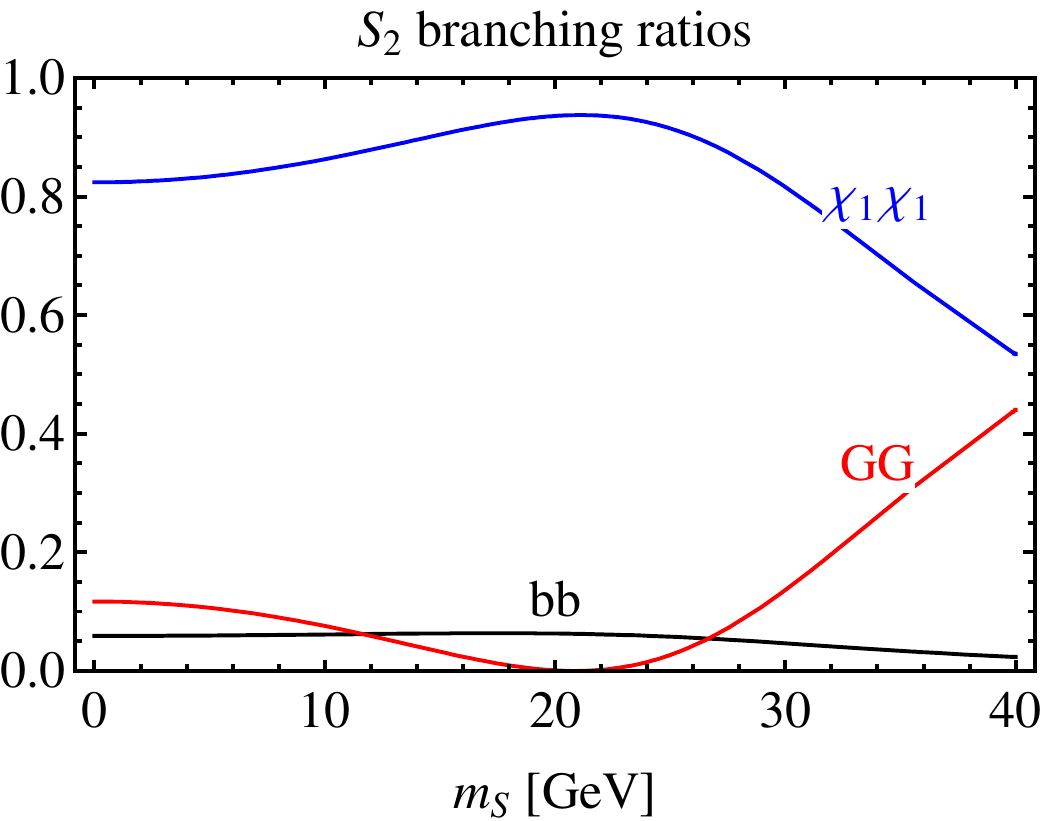}\quad
\includegraphics[width=7cm]{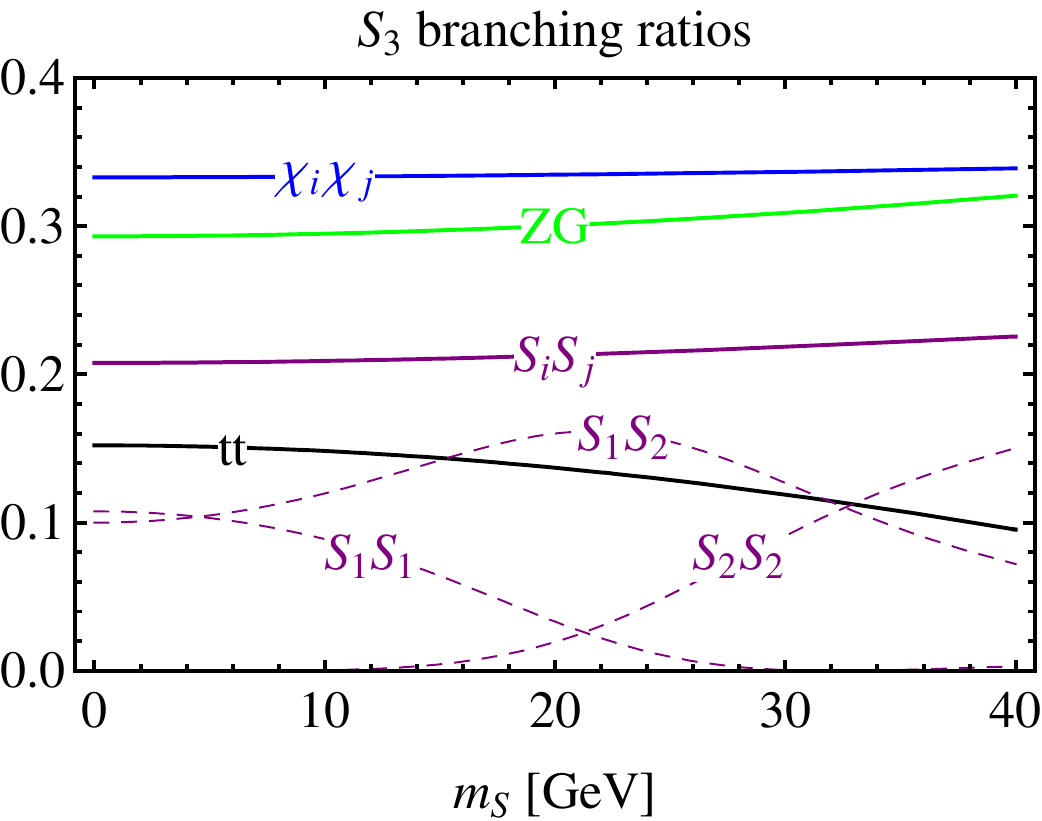}\caption{{\small The dominant branching
ratios of $S_{2}$ (left) and $S_{3}$ (right) for $(\lambda,\tan{\beta
},A_{\lambda})=(0.7,2,400 \mathrm{GeV})$.}}%
\label{BRS3}%
\end{figure}

All the couplings and decay rates for the other Higgs bosons are easily
determined from the parameters of the model as given in the previous Section.
Table \ref{S123} and Figs. \ref{couplS1},\ref{BRS3} illustrate the main
features of the most relevant quantities for $(\lambda,\tan{\beta},A_{\lambda
})=(0.7,2,400~$GeV$)$ and $0<m_{S}\lesssim40$ GeV (see Fig. \ref{paramfinal}).

Using these numbers one can make a preliminary conclusion that observing these
states at the LHC will not be easy, since the production cross sections are
suppressed, and the dominant decay products do not allow for easy background
discrimination. Obviously, a more detailed study is required to assess the LHC
discovery potential.

\begin{table}[ptb]
\centering
\begin{tabular}
[c]{|l|l|l|}\hline
& Production coupling & Branching ratios\\\hline
$S_{1}$ & $\xi_{S_{1}tt},\xi_{S_{1}VV}\lesssim20\%$ (Fig. \ref{couplS1}) &
$BR(GG)$ $\geq98\%)$\\\hline
$S_{2}$ & $\xi_{S_{2}tt},\xi_{S_{2}VV}\simeq100\%$ & $%
\begin{array}
[c]{l}%
\text{See Fig.\ref{BRS3}:}\\
BR(\chi_{1}\chi_{1})=50\div90\%\text{ }\\
BR(GG)\simeq1-BR(\chi_{1}\chi_{1})\text{ }%
\end{array}
$\\\hline
$S_{3}$ & $\xi_{S_{3}tt}\simeq20\%$, $\xi_{S_{3}VV}$ negligible & $%
\begin{array}
[c]{l}%
\text{See Fig.\ref{BRS3}:}\\
BR(\chi_{i}\chi_{j})\simeq35\%\text{ (of which }50\%\text{ into }\chi_{1}%
\chi_{1}\text{)}\\
BR(ZG)\simeq30\%\\
BR(S_{i}S_{j})\simeq20\%
\end{array}
$\\\hline
\end{tabular}
\caption{{\small The neutral CP-even Higgs boson dominant decay modes and the
couplings relevant for their production at the LHC via gluon fusion and vector
boson fusion processes, for $(\lambda,\tan{\beta},A_{\lambda})=(0.7,2,400~$%
GeV$)$ and $0<m_{S}\lesssim40$ GeV.}}%
\label{S123}%
\end{table}

\subsection{Fine tuning}

\label{finetune}

\begin{figure}[ptb]
\centering
\includegraphics[width=8cm]{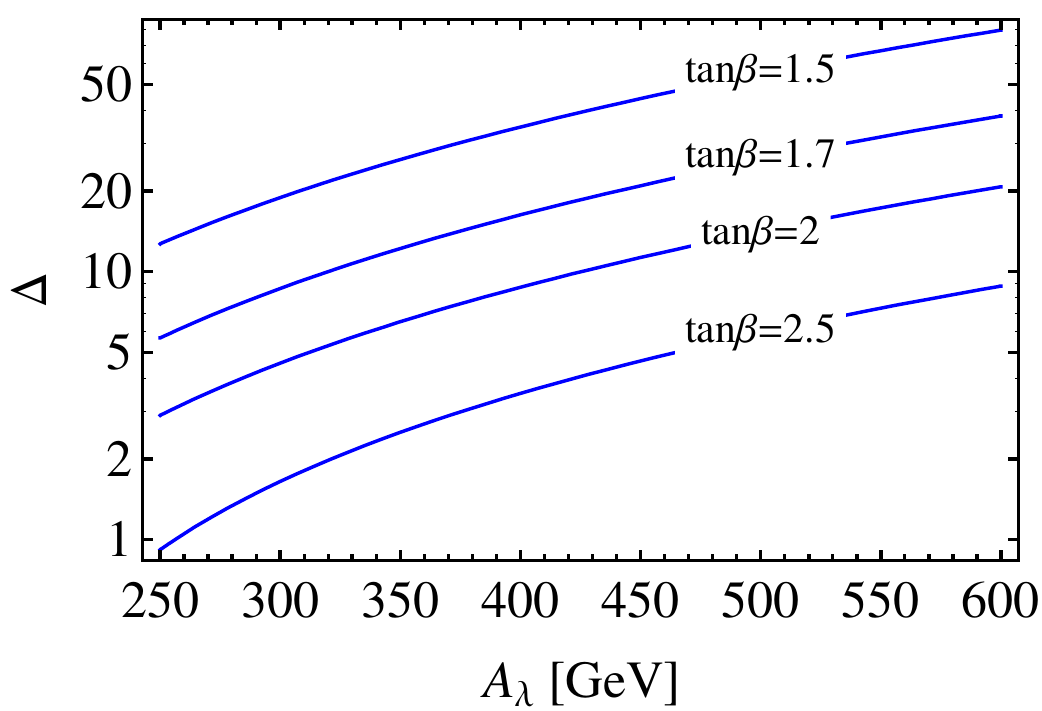} \caption{{\small The logarithmic
derivative Eq. (\ref{logdiv}) plotted as a function of $A_{\lambda}$ for
$(\tan\beta,\lambda)=(1.5,0.65),(1.7,0.7),(2,0.7),(2.5,0.75)$. Smaller
$\tan\beta$ give bigger $\Delta$ (and hence require bigger finetuning). In
this plot we assumed $m_{S}=0$, but the change for small allowed values of
$m_{S}$ is negligible.}}%
\label{delta}%
\end{figure}

It is interesting to know an estimate of the finetuning required to satisfy
the various restrictions on the parameters of the model under consideration.
For an early discussion of finetuning in the NMSSM see \cite{king}. The first
thing to check is if there is too strong dependence of the Z-mass, or of the
vacuum expectation value $v$ in eq. (\ref{lv}), on the various parameters. The
strongest such dependence is on the parameter $A_{\lambda}^{2}$. Naively this
would seem to require finetuning of the order $A_{\lambda}/\lambda^{2}%
v^{2}\sim10$ for $A_{\lambda}=400$ GeV. However, this estimate does not take
into account the fact that the two cancelling terms are not totally
independent: the variation of $A_{\lambda}$ influences the other term via the
angle $\beta$ as determined by the second equation (\ref{sin2b}). An estimate
which takes this effect into account is given by the logarithmic
derivative\footnote{As a consequence of the above effect the derivative
$\partial v^{2}/\partial A^{2}$ is actually \textit{negative}.}%
\begin{equation}
\Delta=\left\vert \frac{\partial\log v^{2}}{\partial\log A^{2}}\right\vert \,,
\label{logdiv}%
\end{equation}
which can be evaluated numerically, see Fig. \ref{delta}. We see that
$\tan\beta$ below $\sim1.7$ starts to be disfavored by this finetuning,
although $\tan\beta=2$, $A_{\lambda}=400$ GeV is OK with less than $10\%$ finetune.

\begin{figure}[ptb]
\centering
\includegraphics[width=8cm]{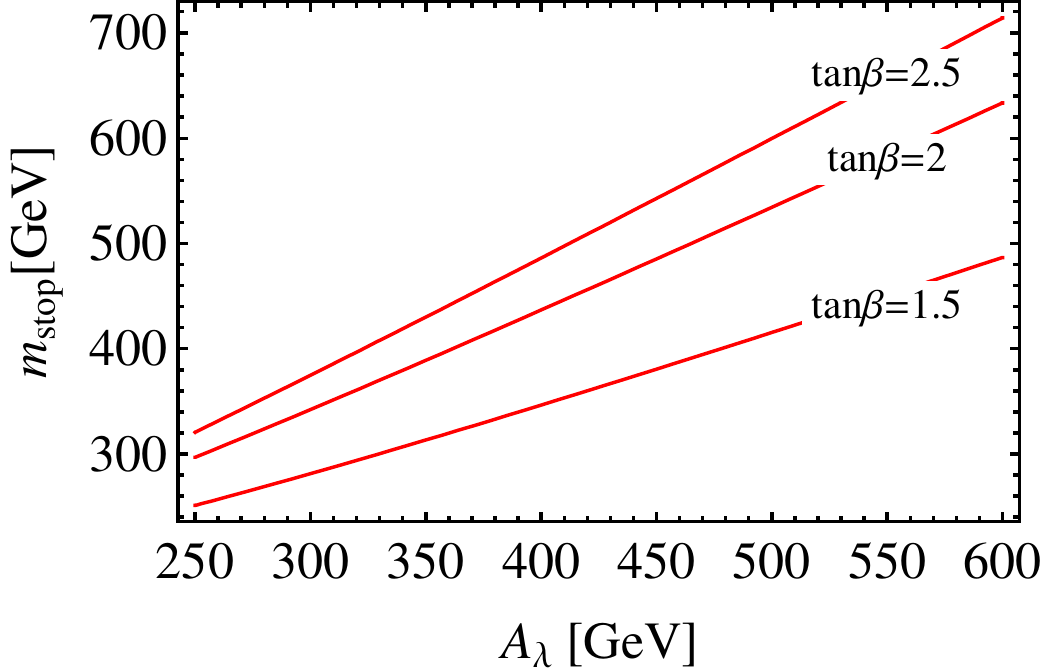} \caption{{\small The maximal value of
physical stop mass (with no mixing) which is consistent with the naturalness
bound $\delta(m_{2}^{2}-m_{1}^{2})/(m_{2}^{2}-m_{1}^{2})<1$, see Eq.
(\ref{deltastop}). From below up: $(\tan\beta,\lambda
)=(1.5,0.65),(2,0.7),(2.5,0.75)$. Mediation scale $\Lambda_{\text{mess}}=100$
TeV. In this plot we assumed $m_{S}=0$, but the change for small allowed
values of $m_{S}$ is negligible.}}%
\label{stopmax}%
\end{figure}

Whereas these considerations apply to the dependence of $v$ on the low energy
parameters, it is also necessary to check the consistency of the values of
these same parameters with the expected contributions due to RGE
evolution\footnote{Alternatively we could look directly at the dependence of
$v$ on the high energy paremeters, which are considered more fundamental.}.
Here the difference $m_{2}^{2}-m_{1}^{2}$ gets a contribution due to the stop
mass:%
\begin{equation}
\delta(m_{2}^{2}-m_{1}^{2})=\frac{3y_{t}^{2}}{4\pi^{2}}(m_{\tilde{t}}%
^{2}-m_{t}^{2})\log\frac{\Lambda_{\text{mess}}}{v}\text{ (no mixing)}
\label{deltastop}%
\end{equation}
where $\log{\Lambda_{\text{mess}}/v}=6\div40$ for $\Lambda_{\text{mess}}%
=100~$TeV$\div10^{19}~$GeV$.$ This equation can be used to set an upper bound
on the stop mass with no finetuning, i.e. for $\delta(m_{2}^{2}-m_{1}%
^{2})/(m_{2}^{2}-m_{1}^{2})=1,$ see Fig\ \ref{stopmax}, where we take low
mediation scale, $\Lambda_{\text{mess}}=100$ TeV\footnote{To make this plot,
parameters $m_{1}^{2}$ and $m_{2}^{2}$ have to be expressed in terms of
$v,\lambda,\tan\beta,A_{\lambda}$ and $m_{S}$ from equations (\ref{lv}),
(\ref{sin2b}).}. We see that $m_{\tilde{t}}=300$ GeV, as assumed above, is
safely within the allowed range.

Finally we notice that, although $v$ in eq. (\ref{lv}) is only weakly
dependent on $m_{S}^{2}$, the current experimental constraints mostly on the
chargino mass (see eq. (\ref{mueff}) and Fig. \ref{param}) require $m_{S}$ to
be below $40\div50$ GeV. At the same time there is a one-loop contribution to
the running of $m_{S}^{2}$ due to $A_{\lambda}^{2}$:%
\begin{equation}
\delta m_{S}^{2}=\frac{1}{4\pi^{2}}\lambda^{2}A_{\lambda}^{2}\log\frac
{\Lambda_{\text{mess}}}{v} \label{mSA}%
\end{equation}
For $\lambda=0.7$, $A_{\lambda}=400$ GeV, and $\Lambda=100$ TeV this gives
$\delta m_{S}^{2}=(110$ GeV$)^{2}$. Since the allowed range of $m_{S}^{2}$ is
a factor $5\div10$ smaller, to comply with this limit it is clear that the
model under consideration would again prefer low mediation scale.

\section{Conclusions and outlook}

\label{conclusions}

Even if we assume, with good reasons indeed, that supersymmetry is relevant in
nature, there is no water-tight argument
that requires the presence of supersymmetric signals at the LHC. Our best hope
is a natural solution of the hierarchy problem of the Fermi scale, which makes
such a presence likely. For this to be the case, however, requires a low level
of fine tuning in the $Z$-mass, i.e. a maximally natural solution of the
hierarchy problem. The LEP limit on the Higgs boson mass is particularly
important, since it excludes the most natural regions of parameter space of
the simplest supersymmetric models. This amply motivates the focus on
supersymmetric extensions of the SM that minimize this fine-tuning and remain,
at the same time, reasonably simple.

A particularly simple possibility for increasing the Higgs mass is to add a
new quartic interaction for the Higgs doublets via the superpotential
interaction $\lambda SH_{1}H_{2}$. However, to maintain perturbative
unification of the gauge couplings, a clear success of weak scale
supersymmetry, the value of $\lambda$ is limited. With minimal matter content
this interaction provides at most an additional $\sim10$ GeV to the Higgs
boson mass, leading to only a small allowed region of parameter space, even
including mixing amongst the Higgs bosons. In contrast, with additional matter
the perturbative evolution of couplings allows a larger value of $\lambda$,
increasing the Higgs boson mass by up to $\sim20$ GeV compared to the theory
without the singlet field. Furthermore, in this case mixing can augment the
Higgs boson mass by another 2--8 GeV, considerably enlarging the allowed
region of parameter space.

The extra matter implies that the gauge couplings are larger in the UV, and
the top coupling smaller, compared to the minimal matter case. Providing the
extra matter fills complete SU(5) multiplets, the successful unification of
gauge couplings at 1 loop is unaltered. The changes from 2 loops and threshold
corrections depends on the nature of the extra matter. In the case of $5 +
\bar{5}$ representations, the prediction for $\alpha_{s}(M_{Z})$ from 2 loop
running is decreased, improving the agreement with data, but this is offset by
an expected increase in $\alpha_{s}(M_{Z})$ from threshold corrections from
non-degeneracies within the $5 + \bar{5}$ multiplets. The situation with $10 +
\bar{10}$ representations is the opposite: 2 loop running increases the
discrepancy with data, but is countered by the threshold corrections. In
either case, the significance of gauge coupling unification is comparable to
the standard case with minimal matter.

In this work we have considered PQ SUSY---a version of the NMSSM that
incorporates
the above mechanism for enlarging the Higgs mass and is fully realistic, with
a minimum number of parameters. The superpotential is assumed to be exactly
invariant under a Peccei-Quinn symmetry, elegantly solving the $\mu$ problem,
while the soft scalar interactions include small PQ breaking interactions to
give a mass to the pseudo-Goldstone boson $G$. The theory possesses just two
parameters more than the MSSM -- one is the soft mass parameter for the scalar
$S$, $m_{S}^{2}$, and the other is the mass for $G$. A combination of vacuum
stability and chargino mass limits implies a restricted range for $m_{S}^{2}$,
so that a $10-20\%$ fine tuning is necessary, and a low messenger scale is preferred.

The Higgs boson system has a few characteristic properties in its spectrum and
in its couplings. The spectrum contains two CP-even neutral scalars relatively
close in mass, one above and one below the \textquotedblleft naive" LEP bound
of 115 GeV by $10\div20$ GeV, and with a shared coupling to the vector boson
pairs, VV. Only the sum of these coupling squared is close to the squared
coupling of the SM Higgs boson to VV. Related to the approximate Peccei-Quinn
symmetry, the Higgs boson spectrum also contains a CP-odd light state, $G$,
present in the main decay modes of all the CP even neutral scalars. $G$ itself
decays to $b\bar{b}$ and $\tau\bar{\tau}$, with branching ratios close to
those of a light SM Higgs boson. Quite clearly, to assess the discovery
potential of such a Higgs boson system at the LHC or TeVatron requires, and
deserves, a detailed examination.

A small region of parameter space where $2m_{\tau}<m_{G}<2m_{b}$ and $G$
decays into $\tau^{+}\tau^{-}$ is also allowed; in this case the model
predicts $BR(\Upsilon\rightarrow\gamma G)$ within reach of the existing B-factories.

The general phenomenology of the model crucially depends on the properties of
the lightest neutralino, which is predicted to have a mass near 50 GeV.
Naturalness considerations suggest a low scale for supersymmetry breaking,
$\sqrt{F}$, so that the gravitino is the LSP and $\chi_{1}$ the next-to-LSP.
In this case, all superpartner production events at colliders will yield at
least two $\chi_{1}$, with each decaying predominantly into a gravitino and a
pseudo-Goldstone $G$, with a width of order $10^{-5}$~eV for $\sqrt{F}%
=100$~TeV. Given the scaling $\Gamma(\chi_{1})\propto1/F^{2}$, $\chi_{1}$
could therefore decay inside the detector if $\sqrt{F}$ is less than of order
1000~TeV.
Pair production of $\chi_{1}$ at LEP2 has a cross section of order 0.1 pb,
leading in this case to events with 4 $b$ jets and missing energy. We do not
know if searches by the LEP experiments would have detected this signal. If
not, the generic superpartner production signal at LHC/TeVatron may include 4
$b$ jets together with the missing energy. The naturalness argument by itself
is not sufficiently tight to prefer $\chi_{1}$ decays inside rather than
outside the detector.

Finally, the consistency of our model with the ElectroWeak Precision Tests
merits further work; in particular, a non-negligible correction to the T
parameter can be induced by values of $\lambda$ in the region of 0.6--0.8.

It is important to note that there are alternative versions of the NMSSM with
extra matter that incorporate both an enhanced Higgs boson mass, perturbative
gauge coupling unification and a solution to the $\mu$ problem. The absence of
mass parameters in the superpotential may be guaranteed by an $R$ symmetry,
that nevertheless allows the interaction $(\kappa/3)S^{3}$ as well as $\lambda
SH_{1}H_{2}$. In order that the $S^{3}$ interaction not substantially reduce
the Higgs mass, the weak scale value of $\kappa$ should be less than about
$0.1$. However, the form of the renormalization group equations allows
$\kappa$ at the unification scale to be close to unity, so this is not a
powerful constraint on the theory. What is the form of the $R$ symmetry
breaking in the supersymmetry breaking scalar interactions? If $A_{\lambda}$
is the only significant $R$ breaking parameter, and $m_{S}^{2}>0$, then this
theory is a perturbation of the model discussed in this paper. On the other
hand there is a new minimum for $m_{S}^{2}<0$ that is very different from the
one examined here, where the $S^{3}$ interaction prevents runaway behavior for
$v_{s}$. There are also models of both the PQ and R types with large values of
the symmetry breaking in the soft scalar interactions, but in these cases
there are several more parameters that enter the phenomenology. Nevertheless,
these models may be of interest since they may remove the need to tune
$m_{S}^{2}$ to small values.

\textbf{Note added.} After completion of this work we became aware of the work
of P. Schuster and N. Toro \cite{schuster} where the NMSSM in the PQ and in
the R-symmetric limits is analyzed with special emphasis on the fine tuning
issue. We believe that the present work usefully complements Ref.
\cite{schuster} in many different aspects.

\section*{Acknowledgements}

We thank Steve King for useful discussions at the early stages of this work,
and in particular for pointing out Ref. \cite{pomarol}. We thank Beate
Heinemann, Michelangelo Mangano, and especially Roberto Tenchini for very
useful discussions of various searches at the TeVatron, LHC, and LEP. We thank
Manuel Drees for pointing out \cite{dreiner}. {The work of R.B. was supported
in part by the EU under RTN contract MRTN-CT-2004-503369, and by the Humbolt
Research Foundation. The work of L.J.H. was supported in part by the NSF grant
PHY-04-57315 and by the US Department of Energy under contract
DE-AC02-05CH11231. }

\appendix

\section{2-loop beta functions}

\label{RGEs}

For the convenience of the reader we give here 2-loop supersymmetric beta
functions used to produce results in Section \ref{maxlambda}. The gauge
coupling beta functions are:
\begin{align*}
\frac{dg_{i}}{d\log\mu}  &  =\frac{1}{16\pi^{2}}b_{i}g_{i}^{3}+\frac{1}%
{(16\pi^{2})^{2}}g_{i}^{3}\left(  \sum_{j=1}^{3}b_{ij}g_{j}^{2}-b_{i;top}%
y_{t}^{2}-b_{i;\lambda}\lambda^{2}\right)  ,\\
b_{i}  &  =\left[
\begin{array}
[c]{c}%
\frac{33}{5}+n_{5}\\
1+n_{5}\\
-3+n_{5}%
\end{array}
\right]  ,\quad b_{i;top}=\left[
\begin{array}
[c]{c}%
\frac{26}{5}\\
6\\
4
\end{array}
\right]  ,\quad b_{i;\lambda}=\left[
\begin{array}
[c]{c}%
\frac{6}{5}\\
2\\
0
\end{array}
\right]  \,,\\
b_{ij}  &  =\left[
\begin{array}
[c]{ccc}%
\frac{199}{25}+\frac{7}{15}n_{5} & \frac{27}{5}+\frac{9}{5}n_{5} & \frac
{88}{5}+\frac{32}{15}n_{5}\\
\frac{9}{5}+\frac{3}{5}n & 25+7n_{5} & 24\\
\frac{11}{5}+\frac{4}{15}n_{5} & 9 & 14+\frac{34}{3}n_{5}%
\end{array}
\right]  \,.
\end{align*}
Our result for $b_{ij}$ agrees with \cite{murayama}. The dependence of
$b_{ij}$ on $n_{5}$ as given in \cite{pomarol} is wrong.

The relevant beta functions of $y_{t}$ and of the NMSSM couplings
$\lambda,\kappa$ are%
\begin{align*}
\frac{dy_{t}}{d\log\mu}  &  =\frac{y_{t}}{16\pi^{2}}\left(  6y_{t}^{2}%
+\lambda^{2}-\frac{13}{15}g_{1}^{2}-3g_{2}^{2}-\frac{16}{3}g_{3}^{2}\right)
-\frac{y_{t}}{(16\pi^{2})^{2}}(22y_{t}^{4}+3\lambda^{2}y_{t}^{2}+3\lambda
^{4}+2\kappa^{2}\lambda^{2})\,,\\
\frac{d\lambda}{d\log\mu}  &  =\frac{\lambda}{16\pi^{2}}\left(  4\lambda
^{2}+3y_{t}^{2}+2\kappa^{2}-g_{1}^{2}-3g_{2}^{2}\right)  -\frac{\lambda
}{(16\pi^{2})^{2}}(10\lambda^{4}+9y_{t}^{2}\lambda^{2}+9y_{t}^{4}+8\kappa
^{4}+12\lambda^{2}\kappa^{2}),\\
\frac{d\kappa}{d\log\mu}  &  =\frac{\kappa}{16\pi^{2}}(6\kappa^{2}%
+6\lambda^{2})-\frac{\kappa}{(16\pi^{2})^{2}}(24\kappa^{4}+24\lambda^{2}%
\kappa^{2}+12\lambda^{4}+18y_{t}^{2}\lambda^{2}\,).
\end{align*}
In our analysis we have omitted the two loop contributions of gauge couplings
to the running of $y_{t}$,$\lambda,\kappa$. This is legitimate since $g_{i}$
do not approach non-perturbative values in the UV.

\end{document}